\documentclass[preprint,aps]{revtex4}
\usepackage{bm}

\begin{document}

\title{Strong asymmetry of microwave absorption by bi-layer conducting ferromagnetic films in the microstrip-line based broadband ferromagnetic resonance}

\author{M.\ Kostylev}
\email{kostylev@cyllene.uwa.edu.au}
\affiliation{School of Physics, The University of Western Australia, 35 Stirling Highway, Crawley WA 6009, AUSTRALIA}

\begin{abstract} 
Peculiarities of ferromagnetic resonance response of conducting magnetic bi-layer films of nanometric thicknesses excited by microstrip microwave transducers have been studied theoretically. Strong asymmetry of the response has been found. Depending on the order of layers with respect to the transducer either the first higher-order standing spin wave mode, or the fundamental mode shows the largest response. 
Film conductivity and lowered symmetry of microwave fields of such transducers are responsible for this behavior. Amplitude of which mode is larger also depends on the driving frequency. This effect is explained as shielding of the asymmetric transducer field by eddy currents in the films. This shielding remains very efficient for films with thicknesses well below the microwave skin depth. This effect may be useful for studying buried magnetic interfaces and should be accounted for in future development of broadband inductive ferromagnetic resonance methods.  
\end{abstract}
\maketitle


\section{Introduction} 

Broadband inductive microwave setups are emerging as a common tool with which to measure dynamical properties of magnetic thin films and nano-structures \cite{Silva, Counil, Crew, self-organized}. Their main part is a section of a coplanar or a microstrip microwave transmission line ("`microwave transducer") on which top the sample under investigation sits. Absorption of microwave power by the transmission line loaded by the sample is measured by a network analyzer \cite{Counil} or by a combination of a  microwave generator and microwave passive components \cite{self-organized}. These setups are useful for accurately measuring spin wave excitations which are used as a probe for characterizing magnetic surfaces and buried interfaces.  However, consideration needs to be made of the experimental setup when analyzing results \cite{Counil, Kim,  Schneider, Patton, Devolder1, Devolder2}. In particular, Schneider et al. \cite{Schneider} mention that measured response is dependent on the distance of the film from the coplanar surface due to a change in strip line impedance. In another work they report on using a floating ground plane above the film in order to enhance the response amplitude \cite{Schneider2}. The broadband inductive technique was also shown to excite not only the fundamental ferromagnetic resonance mode, but also higher order spin wave resonances \cite{Crew}. 

In this paper we present theoretical results showing asymmetric broadband-FMR response of multilayered metallic films with total thicknesses much smaller than the microwave skin depth. This effect is one more feature which distinguishes this technique from the cavity ferromagnetic resonance (FMR). 

On the other hand, recently an interest has emerged to excitation of propagating spin waves in metallic magnetic films by narrow (several microns in width) microstrip microwave transducers \cite{Bailleul, Demidov1, Demidov2, Khitun}. This interest is largely due to the possibility of using spin waves to perform logic operations \cite{Hertel, logic2005, logic2007, nanologic, Khitun2, Souza}. If necessary, the expressions we obtain can be easily extended to describe excitation of propagating spin waves following suggestions in Ref. \cite{nonreciprocity} and Ref. \cite{Kim}.

\section{Dynamic equations} 
To construct our model we use the system of Maxwell equations with no electric-bias term as suggested previously \cite{Wolfram-deWames, Almeida, Kim}:
\begin{eqnarray}
 \nabla \times \textbf{h}& = \frac{4 \pi \sigma}{c}\ \textbf{e}	\\ \nonumber
 \nabla \cdot \textbf{h}& = -\nabla \cdot 4 \pi \textbf{m}	\\  \nonumber
 \nabla \times \textbf{e}& = -i \frac{\omega}{c} (\textbf{h}+4 \pi \textbf{m}) 	
\end{eqnarray}

Here $\textbf{m}$ is the dynamic magnetization, $\textbf{e}$ is an electric field, and $\omega$ is the eigenfrequency of spin waves. To obtain the last of these equations we assumed that all dynamic fields and the dynamic magnetization have the time dependence in the form $exp(i\omega t)$. 

In Ref.\cite{Kim} starting with these equations we constructed a quasi-analytical theory allowing for inhomogeneous dynamic magnetization in the film plane in the form of spin waves with a set of in-plane wave numbers $k_x$. This was important in that work, since due to the type of symmetry in the coplanar transducers, their microwave fields do not excite homogeneous precession  $k_x$=0 \cite{Dmitriev}. On the contrary, microstrip transducers show maximum of microwave absorption at $k_x$=0 (see e.g. \cite{Dmitriev-kalinikos} and references therein) due to symmetry of the in-plane component of the dynamic magnetic field of the transducer $h_x$. For microstrip transducers with a microstrip width much larger than the free spin-wave propagation path (about 100 micron) in Permalloy (which is the metallic magnetic material with the lowest spin wave damping) one can consider the dynamic magnetization and all microwave fields being homogeneous in the film plane (the (x,z) plane). Assuming that the microwave current of the microstrip and the static magnetic field $H$ (which will later enter the magnetic torque equation) are along the axis $z$, these equations
can be reduced to a system as follows:
\begin{eqnarray}
 \label{eq:system}
\partial{h}_x/\partial{y}=-\frac{4\pi \sigma}{c} e_z	\\  \nonumber
\partial{e}_z/\partial{y} = i \frac{\omega}{c} (h_x+4 \pi m_x) 	\\  \nonumber
h_y=-m_y.
\end{eqnarray}

The last of these equations shows that the out-of-plane component of the dynamic magnetic field is purely of magnetostatic nature and represents the dynamic demagnetizing field of $m_y$. It remains present when the sample conductivity is set to zero. On the contrary, from the first two equations one sees that the in-plane component of the magnetic field is of curling nature. Indeed, one can reduce Eqs.(2a) and (2b) to  a second-order differential equation:
\begin{equation}      
-i\delta^2 \partial^2{h_x}/\partial{y}^2+h_x=-4\pi m_x.
\end{equation}
where $\delta=\sqrt{c/(4 \pi \sigma \omega)}$ is the microwave skin depth. 
One sees that in the limit $m_x$=0, when the sources of the magnetostatic (potential) field vanish this equation remains valid. On the other hand, from Eqs.(1) one can find that $h_x$ vanishes for $\sigma=0$. Thus this component is not present in the insulating samples, where all fields are known to be magnetostatic, it is entirely of curling nature, and is due to the eddy current $j_z$ induced in the conducting ferromagnetic films  by the precessing magnetization and the external microwave magnetic field. 

In Ref.\cite{Kim} it has been shown how to obtain a particular solution of such inhomogeneous equation in terms of Green's functions. A general solution of the homogeneous equation has to be added to this solution to obtain the complete solution. This is not difficult to do, if magnetic layers composing the multi-layered film have the same conductivity \cite{Kim's_thesis}. But if interfaces of magnetic metal layers with different values of conductivities are present the final analytic expressions become too cumbersome to be used in computations.    

Therefore, instead of solving Eq.(3) analytically we derive boundary conditions for this equation and solve it numerically together with Landau-Lifshitz-Guilbert equation. 
From the condition of continuity of $h_x$ and $e_z$ at the interface of two magnetic layers $y=L_1$ we find
\begin{eqnarray}
\sigma_2\partial{h_x^{(1)}} / \partial{y}=\sigma_1 \partial{h_x^{(2)}} / \partial{y} \\ \nonumber
h_x^{(1)}=h_x^{(2)},
\end{eqnarray}
where the indices 1 and 2 denote the layer numbers. The boundary conditions at the outer surfaces of the bilayer film are obtained by first allowing the outer space having finite conductivity $\sigma_{out}$. Then the magnetic field in the outer space is described by Eq.(3) with the zero right-hand part. Its solution is obtained requiring vanishing of the microwave magnetic field on both infinities. At one side from the film $y<0$ we have: 
$h_x^{out}=A exp(\sqrt{i} \delta_{out} y)$, and at the other side from the film $y>L_1+L_2$ we have $h_x^{out}=B exp(-\sqrt{i} \delta_{out} y)$. 
Obviously the same boundary conditions (4) are valid at the boundary of a conducting magnetic film with a conducting outer space. Substituting the second of these expressions into Eqs.(4) one obtains a formula which relates  the field \textit{inside} the ferromagnetic layer to its derivative.
\begin{equation}
h_x=-\frac{1}{ \sqrt{i} \delta_{out} } \frac { \sigma_{out} }{ \sigma_2 } \frac{ \partial h_x}{\partial y} \quad \text{at} \quad y=L_1+L_2,  
\end{equation}
where $\sigma_2$ is the conductivity of the second layer of the bi-layer film. To derive the formula which relates $h_x$ to $\partial{h_x} / \partial y$ at the other film boundary we need to include the magnetic field induced by the microstrip. We model the microstrip line as a surface current density $j_z$ at the film surface $y=0$. Then following the same procedure one obtains:
\begin{equation}
h_x=\frac{1}{\sqrt{i} \delta_{out}} \frac{\sigma_{out}}{\sigma_1} \frac{\partial h_x}{\partial y}-j_z \quad \text{at} \quad y=0.  
\end{equation}

In the limit $\sigma_{out} << \sigma_1,\sigma_2$  Eqs.(5) and (6) reduce to:
\begin{eqnarray}
h_x=-j_0 \quad \text{at} \quad y=0,  \\ \nonumber
h_x=0 \quad \text{at} \quad y=L_{1}+L_{2}.
\end{eqnarray}   

(In these expressions we use SI units.)
The second of these formulas shows that the dynamic magnetic field at the film surface not facing the microstrip line is zero. This suggests that total back-reflection of the microwave magnetic field from the boundary between two media with a large difference in electric conductivity takes place. From the second of Eqs.(4) one now finds that the dynamic magnetic field outside the film $y>L_{1}+L_{2}$ vanishes. Thus a conducting magnetic layer with a thickness  much smaller than  $\delta$ efficiently shields the microwave magnetic field. This is in a striking contract with insulating films: the microwave magnetic field of a transducer easily penetrates through magnetic insulators. It equals $-j_z/2$ everywhere for $y>0$ and $j_z/2$ for $y<0$. (see e.g. Eq.(32) in \cite{Dmitriev-kalinikos}). (To obtain this result from our expressions one has to set $\sigma_1=\sigma_2=\sigma_{out}$ and then take the limit of vanishing conductivities.) 
The latter result allows one to consider the total field $h_x$ as a sum of the external field $-j_z/2$ and the field of eddy currents. Then one finds that the field of the eddy currents should grow from $-j_z/2$ at $y=0$ to $j_z/2$ at $y=L_1+L_2$ ensuring no total dynamic magnetic field outside the film  $y<0$ and $y>L_1+L_2$.

The obtained boundary conditions (7) are quite asymmetric. This is a big contrast compared with the cavity resonance where a homogeneous external microwave magnetic field penetrates the sample through both surfaces and the shielding effect is not seen in the boundary conditions (Eq.(4.1) in \cite{Wolfram-deWames}). (We note, that the conditions (7) are in agreement with the theory \cite{Wolfram-deWames} and for the cavity resonances can be transformed into Eqns. (3.3) or (4.1) in the cited work.) 

In this work we numerically solve the linearized Landau-Lifshitz-Gilbert magnetic torque equation \cite{Gurevich} 
\begin{eqnarray}
i\omega \textbf{m}&=& \\ \nonumber
-\gamma [(\textbf{i}_x m_x &+& \textbf{i}_y m_y+M \textbf{i}_z) \\ \nonumber
 \times (\textbf{i}_x (h_x+ 2A/M^2 \frac{\partial^2 m_{x}}{\partial{y^2}}) &+& \textbf{i}_y (h_y+2A/M^2 \frac{\partial^2 {m_{y}}} {\partial{y^2}}) +H_c \textbf{i}_z) ],
\end{eqnarray}
where $M$ is saturation magnetization of a layer, $A$ is the exchange constant for the layer, $H_c=H+i \alpha \omega/\gamma $, $\alpha$ is the Gilbert magnetic damping constant, and $\textbf{i}$ are the unit vectors along the coordinate axes. (The static magnetic field $\bf{H}$ and the equillibrium magnetization are along the $z$-axis.) 
The torque equation is solved together with Eqs.(2c),(3), the electromagnetic boundary conditions Eqs.(7), and the exchange boundary conditions at the layer interface. The latter are cast in the form suggested in \cite{Barnas}. Provided there is no pinning of magnetization in both layers at the interface these exchange boundary conditions at the interface $y=L_1$ of two layers $\it1$ and $\it2$ read
\begin{eqnarray}
 	\partial{m_{x}^{(1)}}/\partial{y} + \frac{A_{12}}{A_1} m_{x}^{(1)} - \frac{A_{12}}{A_1}\frac{M_1}{M_2} m_{x}^{(2)}=0 \\ \nonumber
 	\partial{m_{x}^{(2)}}/\partial{y} + \frac{A_{12}}{A_2} m_{x}^{(2)} + \frac{A_{12}}{A_2}\frac{M_2}{M_1} m_{x}^{(1)}=0,
\end{eqnarray}
where $A_{12}$ is the inter-layer exchange constant. For the out-of-plane dynamic magnetization components the inter-layer boundary conditions are the same. The exchange boundary conditions at the outer surfaces of the film are, as follows \cite{Soohoo} 
\begin{equation}
\partial{m_{x}}/\partial{y} \pm d{m_{x}}=0, \quad \partial{m_{y}}/\partial{y}=0,
\end{equation}
where $d$ is the pinning parameter at the film surface. The positive sign is for the boundary $y=0$ and the negative one is for $y=L_1+L_2$.
To obtain the numerical solution the system of differential equations is transformed into a system of finite-difference equations. The latter form a matrix-vector equation with a band matrix of coefficients. This linear algebraic system is solved  using numerical methods of linear algebra. The way we incorporate boundary conditions between layers and at the outer surfaces of the film into the equations is shown in the Appendix.

\section{Transducer response}
There are two ways to treat the transducer response. One of them is that of the effective microwave susceptibility \cite{Devolder1, Devolder2, Patton}. It assumes that the total microwave energy absorbed by the material is lost due to magnetic losses by driven precession of magnetization in the area of localization of the microwave magnetic field of the transducer. Another method was established several decades ago when magnetostatic spin waves in monocrystalline ferrimagnetic films of yttrium iron garnet (YIG) were found being promising for processing microwave signals \cite{TIIER}. It was suggested that linear impedance could be used to quantitatively characterize efficiency of excitation of propagating spin waves by microstrip (see e.g. \cite{Ganguly, Emtage, Vugalter, Dmitriev-kalinikos, Kolodin} and references therein), and coplanar \cite{Dmitriev} transducers. The linear impedance is calculated as a ratio of the Poynting vector of the flux of energy of microwave field through the transducer surface to the microwave current in the transducer \cite{Dmitriev-kalinikos}. This approach remains valid also in the case when the transducer width is much larger than the spin wave free propagation path and the most of the irradiated energy is not carried away by spin waves but is lost due to relaxation processes within the reach of the microwave Oersted field of the transducer. In this case it naturally incorporates possibility of  energy losses due to irradiation of propagating spin waves which leave the area of the transducer's microwave Oersted field and carry energy away. This additional loss mechanism for metallic magnetic films was recently discussed in Ref.\cite {Counil} and experimentally studied in detail in \cite{Bailleul, Demidov1, Demidov2}.  Both the effective microwave susceptibility and the linear impedance can be related to the scattering coefficient $S21$ measured in the network-analyzer based broadband FMR \cite{Counil}. 

In this work we choose the complex linear impedance $Z_r$ of a microstrip loaded by the ferromagnetic film as a quantity describing the efficiency of microwave absorption. This is because in this way, if necessary, our theory can be easily extended to include the finite width of the transducer to describe effects of irradiation of propagating spin waves with nonvanishing wave numbers by narrow transducers which were used in recent experiments \cite{Bailleul, Demidov1, Demidov2}. Ref. \cite{nonreciprocity} shows the way the irradiation of propagating waves can be treated. 

Our calculations are carried out following the suggestion in \cite{Dmitriev-kalinikos} originally made for magnetostatic spin waves in YIG films. We adopt this approach to metallic ferromagnetic films. Note, that the latter couple to the transducers much more weakly and take much less power from the transducer. For the in-plane homogeneous surface current and the in-plane homogeneous microwave electric field   the expression for $Z_r$ Eq.(27) from \cite{Dmitriev-kalinikos} reduces to: 
\begin{equation}
Z_r=\frac{e_z(y=0)}{j^*_z w},
\end{equation}
where $w$ is the transducer width in the direction $x$. 

 The obtained values of $Z_r$ are then transformed into the value of the scattering coefficient $S21$. We start with the formula for the input impedance $Z_f$ of a section of a microstrip line loaded by a magnetic film. The film sample has a length $l$ along $z$ and sits on top of the transducer. Following Eq.(25) in \cite{Dmitriev-kalinikos} we have
\begin{equation}
Z_f=Z_c \frac{z_0 cosh(\gamma_f l)+ Z_c sinh ((\gamma_f l)) }{z_0 sinh(\gamma_f l)+ Z_c cosh ((\gamma_f l)) }. 
\end{equation}
In our case of the broadband FMR $z_0$ is the characteristic impedance of the sections of the microstrip line not covered by the sample ("unloaded microstrip") and equals 50 Ohms.  
$Z_c$ is the characteristic impedance of the section of the microstrip line loaded by the sample ("loaded microstrip")
\begin{equation}
Z_c=\sqrt{(Z_0+Z_r)/Y_0}
\end{equation}
with $Z_0$ and $Y_0$ being the complex series  resistance and the complex parallel conductance of the unloaded microstrip, and $\gamma_f$ is the complex propagation constant of the loaded microstrip
\begin{equation}
\gamma_f=\sqrt{(Z_0+Z_r)Y_0}.
\end{equation}
The transmission matrix $T$ of the loaded microstrip is obtained following Ref.\cite{Barry}:
\begin{equation}
\mathrm{T} = [\mathrm{T^{(1)}} \cdot \mathrm{T^{(2)}}\cdot
\mathrm{T^{(3)}}], 
\end{equation}
where $T^{(1)}$ and $T^{(3)}$ are the transmission matrices of junctions of the loaded microstrip with the unloaded microstrip. The former is for the junction at the front edge of the sample and the latter is for the rear edge.  
These matrices are defined via the complex reflection coefficient: 
\begin{equation}
\Gamma= \pm \frac{Z_f-z_0}{{Z_f+z_0}},
\end{equation} 
where the positive sign is for the front edge and the negative sign is for the rear edge.
The elements of these matrices are:
\begin{eqnarray}
 \mathrm{T_{11}^{(1,3)}} = \mathrm{T_{22}^{(1,3)}}
 = (1-\Gamma)^\mathrm{-1}, \\ \nonumber
   \mathrm{T_{12}^{(1,3)}} =
\mathrm{T_{21}^{(1,3)}} = \Gamma (1-\Gamma)^\mathrm{-1}.
\end{eqnarray}
The transmission matrix for the loaded microstip between these two edges  $T^{(2)}$ has only diagonal elements:
\begin{equation}
\mathrm{T^{(2)}_{11}} = \mathrm{1 /T^{(2)}_{22}} = e^{ \gamma_f l}.
\end{equation}
The scattering parameter $S21$ of the whole loaded microstrip is $1/T_{22}$ (Eq.(15)). By multiplying the matrices in Eq.(15) one obtains:
\begin{equation}
S21=\frac{\Gamma{^2}-1}{\Gamma{^2}e^{ \gamma_f l}-e^{- \gamma_f l}}.
\end{equation}

The magnitude of the linear radiation impedance of magnetostatic spin waves in YIG is of order of $z_0$ \cite{Dmitriev} and one has to use Eqs.(12-14) as they are. But for nanometric metallic films and wide microstrip transducers $Z_r<<50 \Omega$ which is clearly seen in experiment as $\mid \Gamma \mid <<1$ \cite{ Devolder2, Kostylev-experiment}.  This allows considerable simplification  of these formulas. From the condition 
$Z_r<<z_0$ to the first order in $Z_r/z_0$  from Eqs.(12-20) one obtains
\begin{equation}
S21/S21_0=\exp[-\frac{Z_r}{2z_0}l],
\end{equation}
where $S21_0$ is the scattering parameter of the transducer with no sample on its top. 

\section{Discussion} 
Results of numerical calculation using this formalism for the frequencies $\omega/(2\pi)$ 4, 7.5, and 18 GHz are shown in Fig. 1. 
We consider a Cobalt (Co)-Permalloy (Py) bi-layer film with the parameters shown in the figure caption. The bilayer has a thick Py layer and a thin Co layer which role is to introduce dynamic magnetization pinning \cite{Wigen} of magnetization in the Py layer at the layers' interface while avoiding formation of additional resonances localized in the Co-layer.

From the figure one sees that the absorption when the Co-layer faces the microstrip is a few times smaller than when the Permalloy layer faces it. 
One also sees that the amplitude of the first standing spin wave (the second peak from the right) is visible only for Co facing the transducer. It grows with frequency and becomes larger then the fundamental mode (the most right-hand peak) at higher frequencies. This result agrees well with experiment which will be published elsewhere \cite{Kostylev-experiment}.

The next figure (Fig. 2) explains this strong asymmetry of the film response. Its 
upper panel (Fig. 2(a)) shows distribution of dynamic magnetization $m_x(y)$ across the bi-layer thickness when the surface current is applied from the side of the thin Cobalt layer. One observes partial dynamic pinning of magnetization at the Py-interface with Co which manifests itself as considerably inhomogeneous $m_x(y)$-distribution through the Py-layer for the fundamental mode with the minimum at the layer interface. The first higher-order mode of the stack represents a combination of the fundamental mode of the Co-layer and the 1st SSW of the Py-layer. For the surface current applied at $y=L_1+L_2$ one obtains the mirror image of this panel.  

Figure 2(b) shows the $h_x(y)$-dependence for both cases of film orientation with respect to the microstrip. First one sees that the distribution satisfies the boundary conditions Eq.(7), and in order to satisfy them the magnetic field needs to have a large negative gradient through the film. This field may be thought as a combination of the Oersted field of the microstrip and of the shielding field of the eddy current $j_e$ in the film. 
The former is constant through the film and the space $y>L_1+L_2$ and equals $-j_0/2$.  The latter equals $-j_0/2$ at $y\leq0$, grows through the film to reach $+j_0/2$ at $y=L_1+L_2$, and remains equal to this value for $y>L_1+L_2$ to ensure no dynamic magnetic field behind the film. From this consideration on can infer that the direction of the eddy current is opposite to $\textbf{j}_0$. If one looks at the plot of the electric field 
 (Fig. 2(c)) one finds that the eddy current $j_e=\frac{4\pi \sigma}{c} e_z$ has a phase of $\pi$  and thus is indeed anti-aligned to $\textbf{j}_0$. 
 
From Fig. 2(b) one also sees that $h_x(y)$ is practically a linear function. Indeed, the deviation from the linearity for the long solid line is less than 10 percent at the maximum of deviation. If one looks at Eq.(3) one finds that without the right-hand part this equation has the solution $h_x^c$ satisfying the boundary conditions (7) as follows:
\begin{equation}
h_x^c= A \sinh [\sqrt{i}\delta(y-L_1-L_2)]  
\end{equation}    
(here we neglect the inter-layer boundary).
In the case case of a thin \textit{monolayer} conducting film $L_1<<\delta, \quad L_2=0$  Eq.(22) reduces to a linear dependence $h_x^c= A \sqrt{i}\delta(y-L_1)$, where $A=2{j_0}/{[\exp (2\sqrt{i}\delta L_1)-1]} \approx j_0/(\sqrt{i}\delta L_1)$. The line slope is $j_0$ which is in agreement with our numerical calculations for monolayer films. This suggests that the contribution to the eddy currents from precessing magnetization (i.e. from the particular solution of the inhomogeneous Eq.3) is small and the eddy current in the ferromagnetic film is predominantly directly induced by the microwave magnetic field of the transducer. The total $h_x$ in the film consists then from the transducer Oersted field and the field of the directly excited eddy current with negligible contribution from precessing magnetization.  
 
From the linearized torque equation (8) one finds that the dynamic magnetization is driven by the total field $h_x$. If one neglects the small contribution from the precessing magnetization to the total field, one can consider (8) as an inhomogeneous equation with the right-hand (driving) term in the form of the linear $h_x(y)$ function. This inhomogeneous system of differential equations can be easily solved analytically, but performing this is out of scope of this paper. Here we just mention that if the resonance modes are well-resolved, the resonance amplitude of the $i$-th resonance $r_i$ should be proportional to the overlap integral $I_i=\int_0^{L_1+L_2}m^{[i]}_x(y) \cdot h_x(y)dy$, where $m^{[i]}_x(y)$ is the magnetization profile of the $i$-th eigenmode of the film. The latter is calculated as an eigenfunction of the operator which is obtained from Eq.(8) by setting $\alpha$=0, $h_x$=0. 

Obviously, if the driven resonances are well-resolved (i.e. the resonance linewidths $\Delta H\approx \alpha \omega/\gamma$ are smaller then distances between the neighboring resonances), the $m_x$ distributions in maxima of resonances (Fig 2(a)) are very close to the respective $m^{[i]}_x(y)$ and can be used to discuss $I_i$.     
As one sees from this figure, when the surface current is applied on the Permalloy side of the bi-layer film the maximum of the proper distribution of dynamic magnetization \textit{for the fundamental mode} $m_x^{[1]}(y)$ coincides with the maximum of the total driving field $h_x(y)$ and the value of the overlap integral $I_0$ is maximized. On the contrary, if the current is applied at the Co-surface of the bi-layer the maximum of the driving field coincides with the minimum of the dynamic magnetization. $I_0$ is noticeably smaller, resulting in a much smaller amplitude of the fundamental mode. 

This consideration does not apply to the first standing spin wave of the bi-layer. The $m_x(y)$ distribution for this mode is a quasi-antisymmetric function, but the profile $h_x(y)$ remains the same as for the fundamental mode (Fig. 2(b)) and is characterized by a large anti-symmetric component. As a result the overlap integral $I_1$ does not considerably depend on the side at which the microwave current is applied, and its value is large. Thus, the behavior of resonance amplitudes in Fig.1 is explained not as increase in excitation efficiency of the 1st SSW, but as decrease in efficiency of excitation of the fundamental mode for the specific bi-layer film orientation with respect to the microwave transducer.  

Since eddy currents are involved in this effect, frequency dependence of the amplitudes takes place. Figure 3 shows $S21/S21_0$ in the maximum of resonance for different modes as a function of frequency. Figure 3(a)) shows the absolute values, and Fig. 3(b) shows the relative amplitudes of the standing-wave modes with respect to the fundamental mode. One sees that the first exchange mode for Co facing the microstrip becomes dominant at 7 GHz or so. 

The model also demonstrates more efficient excitation by the microstrip transducers of higher-order modes in conducting monolayer samples than in insulating films (Fig. 4). In the case of insulating films, if the surface spins are unpinned $d=0$ (Eq.(10)) the only mode which couples to the transducer field  is the fundamental mode \cite{Kittel}. From Fig. 4 one sees that in the case of conducting films the first SSW which is characterized by an anti-symmetric profile $m_x^{[2]}$ may provide considerable response. As seen in our simulation, this response is practically frequency independent which is in contrast to the bi-layer films.  

\section{Conclusion}
In this work we theoretically studied peculiarities of ferromagnetic resonance response of conducting magnetic bi-layer films of nanometric thicknesses excited by  wide microstrip lines. We found strong asymmetry of the response. Depending on ordering of layers with respect to the transducer either the first higher-order standing spin wave mode, or the fundamental mode showed the largest response. Amplitude of which mode is larger also depends on the driving frequency. This theory is in a good agreement with an experiment published elsewhere.

This effect is explained as shielding by eddy currents induced in the film. Our results show that for films with thicknesses well below the microwave skin depth this shielding remains very efficient. This finding may be useful for studying buried magnetic interfaces, as it allows more efficient excitation of higher-order standing spin-wave modes carrying information about interface spins.

\section{Acknowledgment}
The author thanks Prof. Robert L. Stamps and Mr. Rhet Magaraggia from the University of Western Australia for fruitful discussions and proof-reading the manuscript text.

Support from the Australian Research Council and the University of Western Australia is gratefully acknowledged.

\section{Appendix: the discrete model}
Here we show how we construct the discrete model and how we incorporate boundary conditions into it. We demonstrate it using the equation (3) for the dynamic magnetic field $h_x$. The discrete version of Eq.(8) with boundary conditions (9) and (10) has a similar form. 

We use a three-point formula for discrete differentiation to obtain the equation as follows:
\begin{equation}      
(h_{x}^{(j+1)}+h_{x}^{(j-1)}-2h_{x}^{(j)})/\Delta^2+i\delta^{-2} h_x^{(j)}+i 4\pi \delta^{-2} m_x^{(j)}=0,
\end{equation}
where $\Delta$ is the mesh step along $y$ and $j$ is the number of a point on the mesh. We locate the points on the mesh such as no point is at the boundary. In particular the first point on the mesh $j=1$ is at $y=\Delta/2$, the last point $j=n$ is at $y=L_1+L_2-\Delta/2$. The points at the interface of two layers have numbers $n_0$ and $n_0+1$ and are situated at $y=L_1-\Delta/2$ and $y=L_1+\Delta/2$ respectively. 
Eq.(23) is valid for  any values of $j$ except for 1, $n$, $n_0$, and $n_0+1$. For these boundary points boundary conditions should be included into the discrete second derivative in Eq.(23). 

We assume that the axis $y$ goes along a horizontal line from the left to the right. Let us first consider the point $j=n_0$ to the left of the interface $y=L_1$ which is located half-way between $n_0$ and $n_0+1$. We denote an auxiliary point which is located on the interface as  $j=n_0+0.5$. The value of magnetic field at this point $h_{x}^{(n_0+0.5)}$ can be obtained by extrapolating $h_x(y)$ dependence beyond the point $n_0$ from the left using the Taylor expansion. 
First one calculates the field at the point $h_{x}^{(n_0-0.5)}$ which is half way between $n_0$ and $n_0-1$. Using the Taylor series one obtains:
\begin{eqnarray}
& & h_{x}^{(n_0-0.5)} \\ \nonumber
&=&h_{x}^{(n_0-1)}+ (\partial h_{x}^{(n_0-1)}/\partial y) (\Delta/2) \\ \nonumber
& + & (\partial^2 h_{x}^{(n_0-1)}/ \partial y^2) (\Delta/2)^2 \\ \nonumber
& \approx & h_{x}^{(n_0-1)}+(h_{x}^{(n_0)}-h_{x}^{(n_0-2)})/4 \\ \nonumber
&+&(h_{x}^{(n_0-2)}+h_{x}^{(n_0)}-2 h_{x}^{(n_0-1)})/8.
\end{eqnarray}
This formula is easily simplified to read
\begin{eqnarray}
& & h_{x}^{(n_0-0.5)}= 3h_{x}^{(n_0)}/8+3h_{x}^{(n_0-1)}/4-h_{x}^{(n_0-2)}/8. 
\end{eqnarray}
Similarly at the point $n_0+1.5$ which is half way between the two first points $n_0+1$ $n_0+2$ to the right from the boundary we obtain 
\begin{eqnarray}
& & h_{x}^{(n_0+1.5)}= 3h_{x}^{(n_0+1)}/8+3h_{x}^{(n_0+2)}/4-h_{x}^{(n_0+3)}/8. 
\end{eqnarray}        

The value of magnetic field at the point at the boundary $j=n_0+0.5$ is $h_x^{(b)}$. The first derivative $\partial h_{x}/\partial y$ enters the upper of boundary conditions (4). To evaluate it at the left from the boundary  we use  the Taylor expansion again 
\begin{eqnarray}
\partial h_{x}^{(b)}/\partial y =\partial h_{x}^{(n_0)}/\partial y +  (\partial^2 h_{x}^{(n_0)}/ \partial y^2) \Delta/2 \\ \nonumber
\approx (3h_{x}^{(b)}-4h_{x}^{(n_0)}+h_{x}^{(n_0-0.5)})/ \Delta.  
\end{eqnarray}  
Similarly, to the right from the boundary one has
\begin{eqnarray}
\partial h_{x}^{(b)}/\partial y = (-3h_{x}^{(b)}+4h_{x}^{(n_0+1)}-h_{x}^{(n_0+1.5)})/ \Delta. 
\end{eqnarray}
Then substituting Eqs.(27) and (28) into the first of the boundary conditions (4) with Eqs. (25) and (26) we obtain
\begin{eqnarray}
h_{x}^{(b)}& =& \frac{(29 h_{x}^{(n_0+1)}-6h_{x}^{(n_0+2)}+h_{x}^{(n_0+3)}) \sigma_1}{24(\sigma_1+\sigma_2)} \\ \nonumber
&+&\frac{(29 h_{x}^{(n_0)}-6h_{x}^{(n_0-1)}+h_{x}^{(n_0-2)}) \sigma_2}{24(\sigma_1+\sigma_2)}. 
\end{eqnarray} 
This allows one to evaluate the second derivative at the points $n_0$ and $n_0+1$. Thus, instead of (23) for the point $n_0$ we have
\begin{equation}      
4(h_{x}^{(b)}+h_{x}^{(n_0-0.5)}-2h_{x}^{(n_0)})/\Delta^2+i\delta^{-2} h_x^{(n_0)}+i 4\pi \delta^{-2} m_x^{(n_0)}=0,
\end{equation}
with $h_{x}^{(b)}$ and  $h_{x}^{(n_0-0.5)}$ defined by Eqs.(29) and (25) respectively.  
A similar expression is easily obtained for $n_0+1$. These finite difference equations now incorporate the inter-layer electro-dynamic boundary conditions (4).
 
Similarly, at the outer boundary $y=L_1+L_2$ from (9) we have $h_{x}^{(b)}=0$.
Then from Eq.(30) with $n_0=n$ one derives  the finite difference equation for $j=n$
\begin{equation}      
(13h_{x}^{(n)}-6h_{x}^{(n-1)}+h_{x}^{(n-2)})/(2\Delta^2)+i\delta^{-2} h_x^{(n)}+i 4\pi \delta^{-2} m_x^{(n)}=0,
\end{equation}
where $h_{x}^{(n-0.5)}$ is obtained from Eq.(25) by setting $n_0=n$. In the same way for the point $j=1$ we have  
\begin{equation}      
(-13h_{x}^{(1)}/2+3h_{x}^{(2)}-h_{x}^{(3)}/2)/(2\Delta^2)+i\delta^{-2} h_x^{(1)}+i 4\pi \delta^{-2} m_x^{(1)}=4\eta j_0,
\end{equation}
where $\eta=1/80$ is the factor which relates $j_0$ measured in Ampere per meter to $h_x$ measured in Oersteds. 

Eq.(8) can be discretized in a similar way incorporating exchange boundary conditions (9) and (10) into the exchange operator $2A/M^2 \frac{\partial^2 \bf{m}}{\partial{y^2}}$ at the points $j=1,n_0,n_0+1,n$. In particular, for the point $n$ one has
\begin{equation}
\frac{\partial^2 {m_x^{(n)}}}{\partial{y^2}}\approx \frac{(d\Delta-2)(6m_x^{(n-1)}-m_x^{(n-2)})-(13d\Delta-10)m_x^{(n)}}{2\Delta^2(3-d\Delta)} ,
\end{equation}    
where $d$ is the surface spin pinning parameter from Eq.(10).  

The derived finite-difference equations form a system of linear algebraic equations $\hat{C} \vec{u}=\vec{f}$, where $\vec{u}$ consists of amplitudes of $m_x$, $m_y$, and $h_x$ at the points $j=1,2...n$. The matrix of coefficients of this system $\hat{C}$ has a band form. The system is inhomogeneous, with just one element of the vector $\vec{f}$ being non-zero.  This element is the right-hand side of Eq.(32). It plays the role of the "excitation term" for the motion of magnetization. 
The system $\hat{C} \vec{u}=\vec{f}$ can be easily solved using numerical methods of linear algebra. 
\

\section{Figure captions}
Fig. 1. (Color online) Transmission coefficient $S21/S21_0$ for a bi-layer ferromagnetic metallic film. Layer 1: thickness: 10 nm, saturation magnetization $4 \pi M$: 15080 Oe, exchange constant: $1.0 \cdot 10^6$ erg/cm, Gilbert damping constant: 0.016, conductivity: $1.8 \cdot 10^7$ Sm/m. Layer 2: thickness: 87 nm, saturation magnetization $4 \pi M$: 8042 Oe, exchange constant: $0.55 \cdot 10^6$ erg/cm, Gilbert damping constant: 0.008, conductivity: $4.5 \cdot 10^6$ Sm/m.  Interlayer exchange constant $A_{12}=2 \cdot 10^6$ erg/cm, gyromagnetic constant: $\gamma=2\pi \cdot 2.92$ rad$\cdot$MHz/Oe, spins at the outer surfaces of the film are unpinned $d$=0. Microwave transducer is a microstrip line 1.5 mm in width. (a) microwave frequency is 4 GHz, (b) 7.5 GHz, (c) 18 GHz. Thick lines: amplitudes (left axes); thin lines: phase (right axes). Solid lines: Co layer facing the transducer, dashed lines: Py layer facing the transducer.

Fig. 2 (Color online) Distributions across the film thickness. (a) in-plane component of dynamic magnetization $m_x$.  
(b) in-plane dynamic magnetic field $h_x$ for the fundamental mode.  (c) microwave electric field $e_z$ for the fundamental mode. In all panels thick lines are for the fundamental mode, thin lines are for the first standing spin wave mode. Black solid lines are for the Co layer facing the transducer. Dashed red line is for the Py layer facing the transducer.  Blue lines: phase (right axes). All the other parameters are as for Fig. 1.

Fig. 3. (Color online) (a) absolute values of absorption amplitudes $S21/S21_0$ as a function of frequency. (b) relative amplitudes of standing spin-wave modes with respect to the amplitude of the fundamental mode. Bold solid lines: fundamental mode, thin blue lines: first standing spin-wave mode. Other thin lines: higher-order standing spin waves seen in Fig. 1(c).  Solid black and blue lines: Co layer facing the transducer. Dashed red lines: Py layer facing the transducer. Parameters of calculation are the same as for Fig. 1.

Fig. 4. (Color online) Absolute values (right axis, black lines) of absorption amplitudes $S21/S21_0$ and their ratio (left axis, dark cyan line) as a function of frequency for a monolayer Permalloy film 97 nm in thickness. Thin line: 1st standing spin-wave mode; thick line: fundamental mode. All the other parameters are as for the Permalloy layer in Fig. 1.

\end{document}